\documentclass[twocolumn,aps,pra,showpacs]{revtex4}

\usepackage{graphicx}
\usepackage{dcolumn}
\usepackage{bm}
\usepackage{amsmath}
\usepackage{amssymb}

\begin{document}

\title{Coherent transients mimicked by two-photon coherent control
of a three-level system}

\author{Jongseok Lim}
\author{Jae-uk Kim}
\author{Sangkyung Lee}
\author{Jaewook Ahn} \email{jwahn@kaist.ac.kr}

\affiliation{Department of Physics, KAIST, Daejeon 305-701, Korea}%

\date{\today}

\begin{abstract}
We show that two-photon coherent control in a $V$-shape three-level
system projects one-photon coherent transient in a simple two-level
system. Higher order chirps of a shaped laser pulse play the roles of time and linear chirp in conventional coherent transients. In a devised scheme of a three-pulse coherent excitation experiment, the phase and amplitude of controlled transition probability is retrieved from a 2D Fourier-transform spectral peak.
\end{abstract}

\pacs{32.80.Qk, 32.80.Wr, 42.50.Md, 42.65.Re}

\maketitle


\section{INTRODUCTION}

Recent advances in ultrafast laser and optical pulse shaping
techniques have brought the use of shaped pulses of optical
frequency for the manipulation of quantum systems~[1-4]. This field,
known as ¡°quantum control¡±, though being started as a theoretical
exercise, has rapidly become an experimental reality in a vast
variety of materials extending from atoms and molecules to condensed
matter and biological materials~[5-9].

One of the simplest way to shape an optical pulse is to chirp, or to make a
quadratic spectral phase, i.e., %
\begin{equation}
\phi(\omega)=\frac{a_2}{2}(\omega-\omega_0)^2,
\end{equation}
where $a_2$ is linear chirp rate and $\omega_0$ is laser center
frequency. Chirped pulses have been used to control molecular
vibrational excitation and fragmentation~\cite{CHELKOWSKI,AMSTRUP},
coherent anti-Stokes Raman Scattering microscopy~\cite{Gaubatz},
molecular alignments~\cite{Karczmarek}, and high harmonic
generation~\cite{Lee}, to list a few. Of particular relevance in the
context of the present paper is the chirped pulse excitation of
atoms in the weak-field interaction regime, also known as coherent
transients (CTs)~\cite{Zamith,Degert,Dudovich}.

If we consider one-photon transition in a two-level system of ground
state $|1\rangle$ and excited state $|2\rangle$, for an optical short pulse of gaussian pulse shape with chirp,  the electric field $E(t)$ is given by
\begin{eqnarray}
E(t)&=&{\mathcal{E}_{o}} \exp\left[{-\frac{t^2}{\tau_c^2}}
-i\left(\omega_0 t+\alpha t^2\right)\right], \label{twoleveltime}
\end{eqnarray}
where $\tau_c=\tau_o\sqrt{1+{a_2^2}/{\tau_o^4}}$,
$\alpha=2a_2/(\tau_0^4+4a_2^2)$, and $\tau_o$ is the unchirped pulse
duration. Then the excitation probability amplitude $c_{21}$ at a
finite time $t$ is given in the weak-field regime
as~\cite{Zamith}
\begin{eqnarray}
c_{21}(t)&=&\frac{i\mu_{21}\mathcal{E}_{o}}{\hbar}{}
\int_{t_{0}}^{t}
\exp({-\frac{t^2}{\tau_c^2}}) \nonumber \\
&\times& \exp\left[-i\left((\omega_0-\omega_{21})t'+\alpha
t'^2\right)\right]dt',
\end{eqnarray}
where the finite time integration with a quadratic temporal phase
$\alpha t'^2$ leads to the transient excited-state population being
of a Cornu spiral shape, well known from Fresnel diffraction pattern
from a sharp edge~\cite{Hecht}.

For a short pulse which has broad spectral components, putting chirp
on the pulse delays some of those components with respect to others
in the time domain, and the instantaneous laser frequency shifts as
a function of time. So, from the time when the resonant condition is met, further off-resonant excitation interferes, with the resonant transition, either constructively or destructively, and shows rather an oscillatory
transient behavior. The quantum
interference between resonant and non-resonant excitation
contributions can be easily understood in the frequency domain representation of CTs, given by~\cite{Dudovich2}
\begin{eqnarray}
c_{21}(t)&=&\frac{\mu_{21}}{\hbar} \Big[i\pi
\widetilde{E}(\omega_{21}) \nonumber \\
&+& \wp\int_{-\infty}^{\infty}\frac{\widetilde{E}(\omega)\exp[i(\omega_{21}-\omega)t]}
{\omega_{21}-\omega}d\omega \Big], \label{detwolevel}
\end{eqnarray}
where $\widetilde{E}(\omega)=E_{o}\exp[-{(\omega-\omega_0)^2
\tau_0^2}/{4} +i\phi(\omega)]$ and $\wp$ is the Cauchy principal
value. The time $t$ and the linear chirp rate $a_2$ are the
two control parameters for the quadratic and cubic phase terms, respectively. By varying these two parameters, CTs have found
many coherent control examples, including quantum state holographic
measurements~\cite{Monmayrant}, the coherent transient
enhancement~\cite{Dudovich2}, the time-domain Fresnel
lens~\cite{Degert}, and coherent controls of multi-state
ladders~\cite{Merkel}.

In this paper, we consider a two-photon control of a three-level
system, with a chirped pulse of up to cubic phase, i.e.,
\begin{eqnarray}
\phi(\omega) = a_1(\omega-\omega_0)+\frac{a_2}{2}
(\omega-\omega_0)^2 +\frac{a_3}{6}(\omega-\omega_0)^3,
\label{spectralphase}
\end{eqnarray}
We show in a $V$-type three-level system the given two-photon
coherent control is reduced to the discussed CT in a two-level
system, where now $a_2$ and $a_3$ play the roles of the time and
linear chirp in a regular CT. Furthermore, we apply this coherent
control to a three-pulse coherent control scheme for two-dimensional
Fourier-transform spectroscopy~\cite{Keusters} and measure the
amplitude and phase of two-photon inter-excited states transition
coefficients.

\section{Two-photon coherent control of $V$-shape three-level system}

We consider a three-level system in $V$-type configuration, composed
of ground state $|g\rangle$ and two excited states $|a\rangle$ and
$|b\rangle$, with respective energies $\hbar \omega_g$, $\hbar
\omega_{a}$, and $\hbar \omega_{b}$. The excited states are
dipole-coupled to the common ground state, with dipole moments
$\mu_{ag}$ and $\mu_{bg}$, and the transition between the excited
states is forbidden (i.e., $\mu_{ab}=0$).  Then, the Hamiltonian is
given by
\begin{eqnarray}
H(t)=H_0+V(t),
\end{eqnarray}
where $H_0= \sum_{i}\hbar \omega_{i}|i\rangle\langle i|$ and
$V(t)=-\sum_{i,j}\mu_{ij}E(t)|i\rangle\langle j|$ for $i,j \in \{g,
a, b\}$. With $T=\exp(-i H_0 t/\hbar)$, we transform to the
interaction picture obtaining the interaction Hamiltonian
$H_I(t)=T^{\dag} H(t) T+i \hbar
\frac{dT^{\dag}}{dt}T=\sum_{i,j}V_{ij}(t)
e^{i\omega_{ij}t}|i\rangle\langle j|$, where $V_{ij}(t)=\langle
i|V(t)|j\rangle$ and $\omega_{ij}=\omega_i-\omega_j$.

Then, the transition probability amplitude from state $|i\rangle$ to
state $|f\rangle$, defined by $c_{fi}(t)=\langle f|U_{I}(t,t_{0})|i \rangle$, where $U_{I}(t,t_{0})$ is the evolution operator given by
$U_{I}(t,t_{0})=1-\frac{i}{\hbar}\int^{t}_{t_{0}}H_{I}(t')U_{I}(t',t_{0})dt'$,
is obtained by the order of $V(t)$ as
\begin{eqnarray}
c_{fi}^{(0)}(t)&=&\delta_{fi},  \\
\label{firstorder}
c_{fi}^{(1)}(t)&=&-\frac{i}{\hbar}\int_{t_{0}}^{t}dt'
V_{fi}(t')\exp({i\omega_{fi}t'}),  \\
c_{fi}^{(2)}(t)&=&(-\frac{i}{\hbar})^{2}\Sigma_{j}
\int_{t_{0}}^{t}dt'\int_{t_{0}}^{t'}dt'' \nonumber \\
&\times& V_{fj}(t') V_{ji}(t'')\exp(i\omega _{fj} t'+{i\omega _{ji}
t''}). \label{secondorder}
\end{eqnarray}

For an electric field shaped in frequency domain as
\begin{eqnarray}
\widetilde{E}(\omega)=A(\omega)e^{i\phi(\omega)},
\end{eqnarray}
where $A(\omega)$ is spectral amplitude and $\phi(\omega)$ is
spectral phase, the time domain pulse profile is given by
\begin{align}
\label{efield} E(t)=\frac{1}{\sqrt{2\pi}}\int_{-\infty}^{\infty}
\widetilde{E}(\omega)e^{-i\omega t}d\omega,
\end{align}
where the imaginary part of $E(t)$ is maintained zero all the time
by defining $\widetilde{E}(-\omega)=\widetilde{E}^*(\omega)$ for
$\omega<0$. Then, $V_{ij}(t)=-\mu_{ij}E(t)$, and the first order
(one-photon) transition probability amplitude is simply the corresponding spectral amplitude, given by
\begin{eqnarray}
c_{fi}^{(1)}=i \frac{\mu_{fi}}{\hbar} \sqrt{2\pi}
\widetilde{E}(\omega_{fi}),
\end{eqnarray}
where we consider the integral limit $[t_{0},t] \rightarrow
[-\infty,\infty]$ by assuming the pulse duration is considerably
shorter than all lifetimes involved.

Now we consider two-photon transition probability amplitude between
the excited states $|a\rangle$ and $|b\rangle$. From
Eq.~\eqref{secondorder}, the second order transition probability
amplitude is obtained as~\cite{Dudovich}
\begin{eqnarray}
c_{ba}^{(2)}&=&i\frac{\mu_{ga}\mu_{gb}}{\hbar^{2}}\Big[i\pi
\widetilde{E}^{*}(\omega_{ag})\widetilde{E}(\omega_{bg}) \nonumber \\
&-&\wp \int
\frac{\widetilde{E}^{*}(\omega)\widetilde{E}(\omega_{ba}+\omega)}
{\omega_{ag}-\omega}
d\omega\Big]. \label{eq13}
\end{eqnarray}
For a Gaussian pulse spectrally centered at $\omega_0$,
i.e.,
$A(\omega)=E_0\exp[-{(\omega-\omega_0)^2}/{{\Delta\omega}^2}]$, with
the spectral phase $\phi(\omega)$ given in Eq.~\eqref{spectralphase},
the transition probability amplitude Eq.~\eqref{eq13} is simplified to
the following form
\begin{eqnarray}
c_{ba}^{(2)} = i\frac{\widetilde{\mu}_{ba}}{\hbar^2} \big[i\pi
\widetilde{\mathbb{E}}(\overline{\omega}) -\wp
\int^\infty_{-\infty}\frac{
\widetilde{\mathbb{E}}(\omega)}{\overline{\omega}-\omega} d\omega
\big], \label{reducedVtransition}
\end{eqnarray}
where $\widetilde{\mu}_{ba} =
\mu_{ga}\mu_{bg}\exp[-{\omega^2_{ba}}/{2\Delta
\omega^2}+i{a_{3}}\omega^3_{ba}/24]$, $\overline{\omega} =
({\omega_{ag}+\omega_{bg}})/{2}$, and
\begin{eqnarray}
\widetilde{\mathbb{E}}(\omega) = E^{2}_o
\exp[-2\frac{(\omega-\omega_0)^2}{\Delta\omega^2}
+i\omega_{ba}\dot{\phi}(\omega)].
\end{eqnarray}
It is noted that Eq.~\eqref{reducedVtransition} is of a similar functional form to Eq.~\eqref{detwolevel}, the one-photon transition in a two-level system, except for the sign between the resonant and non-resonant contributions.

\section{Projection to one-photon transition in a two-level system}

The difference between Eq.~\eqref{detwolevel} and Eq.~\eqref{reducedVtransition} is resolved by considering the de-excitation. For an electric field of gaussian pulse shape with
linear chirp (only), the one-photon transition probability amplitude from $|2\rangle$ to $|1\rangle$ is easily found as
\begin{eqnarray}
c_{12}^{(1)}(t)&=&\frac{\mu_{21}}{\hbar}e^{-i(\omega_{21}-\omega_0)t}\Big[i\pi
E^{*}(\omega_{21})e^{i(\omega_{21}-\omega_0)t} \nonumber \\
&-&\wp\int_{-\infty}^{\infty}\frac{E^{*}(\omega)e^{i(\omega-\omega_0)t}}
{\omega_{21}-\omega}d\omega \Big]. \label{detwolevel2}
\end{eqnarray}
As evident from the same structure, the two-photon inter-excited
states transition in a $V$-type system projects one-photon
transition (de-excitation) in a simple two-level system. Tantalizing
part is that, since $\widetilde{\mathbb{E}}(\omega)$ has
differentiated phase, linear chirp in $V$-shape system corresponds to
time in two-level system, and minus quadratic chirp to linear chirp.
Therefore, the obtained solution in Eq.~\eqref{reducedVtransition},
which is the transition probability amplitude $c_{ba}^{(2)}$ for the
two-photon inter-excited state transition in a $V$-type system, has
become formally a one-photon transition probability amplitude, more
specifically a de-excitation process, in a two-level system of
energies 0 and $\overline{\omega}$, induced by the newly defined
electric field $\widetilde{\mathbb{E}}(\omega)$.

Therefore, if we consider the interaction of the $V$-shape system with
laser shaped pulse of linear and quadratic chirps, then we can
achieve duplicated results of coherent transients in a
two-level system interacting with a linearly chirped pulse. With
this information, we can derive the same form of
Eq.~\eqref{twoleveltime} in $V$-shape system. For this, the electric
field, $\mathbb{E}(t)$ is the inverse Fourier transformation of the
complex conjugate of electric field in frequency domain
$\widetilde{\mathbb{E}}(\omega)$ having $t=\omega_{ba}a_2$ and $a_2=-\omega_{ba}a_3$. Then, the
``CT-like'' transition probability amplitude in $V$-shape system becomes
\begin{widetext}
\begin{eqnarray}
c_{ba}^{(2)}(a_2, a_3)=-\frac{\widetilde{\mu}_{ba}}{\hbar^2}
E^{2}_{o} \frac{\Delta\omega}{
\sqrt{{\widetilde{\tau}_{c}}/{\widetilde{\tau}_0}}}
\exp\left(\frac{i}{2}\tan^{-1}
\frac{2\widetilde{a}_2}{\widetilde{\tau}_{o}^2}+i\theta\right)
\int_{-\infty}^{\widetilde{t}}dt'
\exp\left[-\frac{t'^2}{\widetilde{\tau}_c^2}-i\left((\omega_{ba}-\omega_0)
t'
-\widetilde{\alpha}t'^2\right)\right],
\end{eqnarray}
\end{widetext} where $\widetilde{t}=\omega_{ba}a_2+t$,
$\widetilde{a}_2=-\omega_{ba}a_3$,
$\widetilde{\tau}_o={2\sqrt{2}}/{\Delta\omega}$,
$\widetilde{\tau}_c=\widetilde{\tau_o}
\sqrt{1+{\widetilde{a}_2^2}/{\widetilde{\tau}_o^4}}$,
$\theta=-(\omega_{ba}-\omega_0)t$, and
$\widetilde{\alpha}={2\widetilde{a}_2}/
({\widetilde{\tau}_o^4+4\widetilde{a}_2^2})$.

\section{Three-pulse excitation scheme}

The two-photon control in the previous section can be verified by
measuring the phase and amplitude of $c_{ba}^{(2)}(a_2, a_3)$. This
is achieved by using a three-pulse excitation scheme in a
two-dimensional Fourier-transform spectroscopy.
The second pulse is the control pulse which induces two-photon inter-excited states transition from $|a\rangle$ to $|b\rangle$. For this, the atoms need to be excited to $|a\rangle$ by a pre-pulse. In addition, a third pulse is necessary to measure $c_{ba}^{(2)}(a_2, a_3)$. The measurement of the phase and amplitude of $c_{ba}^{(2)}(a_2, a_3)$ via three-pulse coherent excitation is explained in the following.

As our starting point, the quantum system is in the ground state, i.e.,
$|\psi(t=0-)\rangle = |g\rangle$. By assuming the interaction in the
weak field regime (i.e., $V_{ij}(t)\ll {\hbar}$ for all $t$), we
neglect the higher order terms of $V(t)$ and consider the lowest
order terms of each transitions. Then, the evolution operator for
the first pulse is written in terms of three states $\{|g\rangle,
|a\rangle, |b\rangle \}$ as
\begin{eqnarray}
U_{I}(\alpha)=\left(\begin{array}{ccc} 1
& \alpha_{ga}^{(1)} & \alpha_{gb}^{(1)}\\
\alpha_{ag}^{(1)} & 1 & \alpha_{ab}^{(2)}\\
\alpha_{bg}^{(1)} & \alpha_{ba}^{(2)} & 1
\end{array} \right),
\end{eqnarray}
where $\alpha^{(1,2)}$ denote the transition probability amplitudes,
respectively defined in Eqs.~\eqref{firstorder} and
\eqref{secondorder}, for the first pulse. (Likewise, $\beta^{(1,2)}$
and $\gamma^{(1,2)}$ denote the ones for the second and third
pulses, in the following.) We note that
$\alpha_{ga}^{(1)}=\alpha_{ag}^{(1)*}$ but
$\alpha_{ba}^{(2)}\neq\alpha_{ab}^{(2)*}$. Then, the wave function
after the first interaction is given by $
|\psi(0{+})\rangle=|g\rangle+\alpha_{ag}^{(1)}|a\rangle+\alpha_{bg}^{(1)}|b\rangle.
$ For the second pulse, the time delay $\tau_1$ causes an overall
phase shift of $\exp{[i(\omega _{ng} -\omega_{0})\tau_{1}]}$ to the
$V_{ij}(t)\exp({i\omega_{ij}})$ term in Eqs.~\eqref{firstorder} and
\eqref{secondorder}, relative to the ones for $\alpha^{(1,2)}$.
Therefore, the first and second order transition probability
amplitudes for the second pulse, including the phase shift from the
time delay, are obtained, respectively, $\beta_{fi}^{(1)}\exp[{i(\omega
_{fi} -\omega_{0})\tau_{1}}]$ and $\beta_{fi}^{(2)}\exp{[i(\omega _{fg}
-\omega_{0})\tau_{1}-i(\omega _{ig} -\omega_{0})\tau_{1}]}$, where
the rotating wave approximation is used for $\beta_{ba}^{(2)}$.
Accordingly, the evolution operator for the second pulse, including
the time delay effect, is given by
\begin{widetext}
\begin{eqnarray}
U_{I}(\beta)=\left(\begin{array}{ccc} 1 &
\beta_{ag}^{(1)*}e^{-i\Delta\omega_{ag}\tau_{1}}
& \beta_{bg}^{(1)*}e^{-i\Delta\omega_{bg}\tau_{1}}\\
\beta_{ag}^{(1)}e^{i\Delta\omega_{ag}\tau_{1}}
& 1 & \beta_{ab}^{(2)}e^{i(\Delta\omega_{ag}-\Delta\omega_{bg})\tau_{1}}\\
\beta_{bg}^{(1)}e^{i\Delta\omega_{bg}\tau_{1}} & \beta_{ba}^{(2)}
e^{-i(\Delta\omega_{ag}-\Delta\omega_{bg})\tau_{1}} & 1
\end{array} \right),
\end{eqnarray}
where $\Delta\omega_{ij}=\omega_{ij}-\omega_0$.  Then, after the
second pulse, the wave function becomes $ |\psi{(\tau_{1})}\rangle =
U_{I}(\beta)|\psi(0{+})\rangle$.
Likewise, the evolution operator for the third pulse, including the
effect of the time delay $\tau_2$ relative to the second pulse, is
given by
\begin{eqnarray}
U_{I}(\gamma)=\left(\begin{array}{ccc} 1 &
\gamma_{ag}^{(1)*}e^{-i\Delta\omega_{ag}(\tau_{1}+\tau_{2})}
& \gamma_{bg}^{(1)*}e^{-i\Delta\omega_{bg}(\tau_{1}+\tau_{2})}\\
\gamma_{ag}^{(1)}e^{i\Delta\omega_{ag}(\tau_{1}+\tau_{2})}
& 1 & \gamma_{ab}^{(2)}e^{i(\Delta\omega_{ag}-\Delta\omega_{bg})(\tau_{1}+\tau_{2})}\\
\gamma_{bg}^{(1)}e^{i\Delta\omega_{bg}(\tau_{1}+\tau_{2})} &
\gamma_{ba}^{(2)}e^{-i(\Delta\omega_{ag}-\Delta\omega_{bg})(\tau_{1}+\tau_{2})}
& 1
\end{array} \right),
\end{eqnarray}
\end{widetext}
and, after the all three pulsed interactions, the final wave
function $|\psi{(\tau_{1}+\tau_{2})}\rangle$ is obtained as the sum
of twenty seven different terms. By measuring the projection to
$|b\rangle$ state, the probability $P_b=|\langle b|\psi\rangle|^2$
is given by
\begin{align}
P_{b}(\tau_1,\tau_2)=|{\alpha_{bg}^{(1)}}|^{2}+|{\beta_{bg}^{(1)}}|^{2}
+|{\gamma_{bg}^{(1)}}|^{2}+\dots \nonumber \\
+\alpha_{ag}^{(1)*}\beta_{ba}^{(2)*}\gamma_{bg}^{(1)}e^{i(\Delta\omega_{ag}\tau_{1}
+\Delta\omega_{bg}\tau_{2})}+\dots,
\end{align}
where, for example, the term
$\alpha_{ag}^{(1)*}\beta_{ba}^{(2)*}\gamma_{bg}^{(1)}\exp{(i\Delta\omega_{ag}\tau_{1}
+i\Delta\omega_{bg}\tau_{2})}$ denotes the quantum interference
between the two transitions $|g\rangle \rightarrow |a\rangle
\rightarrow |b\rangle$ and $|g\rangle \rightarrow |b\rangle$. The
coefficient $\alpha_{ag}^{(1)*}\beta_{ba}^{(2)*}\gamma_{bg}^{(1)}$
is retrieved from $|\langle b|\psi\rangle|^2$, as the amplitude and
phase of the temporally modulated component with the function
$\exp{(i\Delta\omega_{ag}\tau_{1} +i\Delta\omega_{bg}\tau_{2})}$. The
modulation $\exp({i\Delta\omega_{ag}\tau_{1}})$ and
$\exp({i\Delta\omega_{bg}\tau_{2}})$ are from the phase evolution that the atoms
are respectively in state $|a\rangle$ during $\tau_{1}$ and in state
$|b\rangle$ during $\tau_{2}$. The 2D Fourier-transform spectrum is defined as
\begin{eqnarray}
S(\omega_1,\omega_2)=\int \int
P_b(\tau_1,\tau_2)e^{-i(\omega_1\tau_1+\omega_2\tau_2)}d\tau_1
d\tau_2,
\end{eqnarray}
which has 49 peaks including a zero frequency peak~\cite{Lim2}. The
coefficients of the spectral peaks of $S(\omega_1,\omega_2)$ in the
first quadrant of the two-dimensional plane are listed in
TABLE~\ref{peakcoefficient}. Aside from the constant
$\alpha_{ag}^{(1)*}\gamma_{bg}^{(1)}$, the controlled transition
probability amplitude $\beta_{ba}^{(2)*}$ is then retrieved from the peak
located at ($\omega_1$, $\omega_2$) = ($\Delta\omega_{ag}$,
$\Delta\omega_{bg}$). As a result, the three-pulse coherent control scheme devised for 2D Fourier-transform spectroscopy can be used to measure the two-photon inter-excited states transition coefficients.
\begin{widetext}
\begin{table*}
\begin{center}
\begin{tabular}{c| c c c}
\hline \hline $\omega_2$  $\backslash$ $\omega_1$ & $\Delta\omega_{ag}$ & $\Delta\omega_{bg}
-\Delta\omega_{ag}$
& $\Delta\omega_{bg}$ \\
\hline  $\Delta\omega_{bg}$&
$\alpha_{ag}^{(1)*}\beta_{ba}^{(2)*}\gamma_{bg}^{(1)}$ &
$\alpha_{ag}^{(1)}\alpha_{bg}^{(1)*}\beta_{ag}^{(1)*}\gamma_{bg}^{(1)}$
& $\alpha_{bg}^{(1)*}\gamma_{bg}^{(1)}$ \\
$\Delta\omega_{bg} -\Delta\omega_{ag}$ &
$\alpha_{ag}^{(1)*}\beta_{ag}^{(1)}\beta_{ba}^{(2)*}\gamma_{ba}^{(2)}$
& $\alpha_{ag}^{(1)}\alpha_{bg}^{(1)*}\gamma_{ba}^{(2)}$
& $\alpha_{bg}^{(1)*}\beta_{ag}^{(1)}\gamma_{ba}^{(2)}$ \\
$\Delta\omega_{ag}$ &
$\alpha_{ag}^{(1)*}\gamma_{bg}^{(1)}\gamma_{ba}^{(2)*}$ &
$\alpha_{ag}^{(1)}\alpha_{bg}^{(1)*}\beta_{ag}^{(1)*}
\beta_{ab}^{(2)*}\gamma_{bg}^{(1)}\gamma_{ba}^{(2)*}$
& $\alpha_{bg}^{(1)*}\beta_{ab}^{(2)*}\gamma_{bg}^{(1)}\gamma_{ba}^{(2)*}$\\
\hline \hline
\end{tabular}
\end{center}
\caption{Peaks in the first quadrant of 2D Fourier
transform plane of $|\langle b|\psi\rangle|^2$. For example, the peak at ($\Delta\omega_{ag}$, $\Delta\omega_{bg}$) represents the quantum interference between $|g\rangle \rightarrow |a\rangle \rightarrow |b\rangle$ and $|g\rangle \rightarrow |b\rangle$ transitions.}
\label{peakcoefficient}
\end{table*}
\end{widetext}

\section{CONCLUSION}
In summary, we have shown that two-photon coherent control in a
$V$-shape three-level system behaves formally like a coherent
transient signal in a two-level system, where the roles of time and
linear chirp in the latter are duplicated by linear and quadratic
chirp rates in the former. For the measurement, a three-pulse excitation scheme is devised, and the phase and amplitude of the controlled transition probability is retrieved from a 2D Fourier-transform spectral peak. It is hoped that this control scheme may harness coherent control capability on 2D Fourier-transform spectroscopy.

\section{acknowledgements}
This research was supported by Basic Science Research Program
through the National Research Foundation of Korea (NRF) funded by
the Ministry of Education, Science and Technology (No.
2009-0090843).

\end{document}